\input{aipcheck}

\documentclass[
    ,final            
  ]{aipproc}
\usepackage[english]{babel}
\usepackage{amsmath,amssymb,latexsym}
\usepackage{epsf,epsfig}
\usepackage{graphics,psfrag,float}
\layoutstyle{6x9}

\newcommand{\vk}{{\mathbf{k}}}

\begin{document}

\title{Chiral effective field theory for nuclear matter}


\author{A.~Lacour}{
  address={Helmholtz-Institut f\"ur Strahlen- und Kernphysik (Theorie) and Bethe Center for Theoretical Physics, Universit\"at Bonn, D-53115 Bonn, Germany}
}

\author{U.-G.~Mei{\ss}ner}{
  address={Helmholtz-Institut f\"ur Strahlen- und Kernphysik (Theorie) and Bethe Center for Theoretical Physics, Universit\"at Bonn, D-53115 Bonn, Germany}
  ,altaddress={Institut f\"ur Kernphysik, Institute for Advanced Simulation and J\"ulich Center for Hadron Physics, Forschungszentrum J\"ulich, D-52425 J\"ulich, Germany} 
}

\author{J.~A.~Oller}{
  address={Departamento de F\'{\i}sica, Universidad de Murcia, E-30071 Murcia, Spain}
}

\begin{abstract}
We  report on the recent developments of a new effective field theory for nuclear matter \cite{LacourA_Oller:2009zt,LacourA_Lacour:2009ej,LacourA_Lacour:qc}.
We  present first the nuclear matter chiral power counting
that takes into account both short-- and long--range inter-nucleon interactions.
It also identifies non-perturbative strings of diagrams, related to the iteration of nucleon-nucleon interactions, which have to be re-summed.
The methods of unitary chiral perturbation theory has been shown to be a useful tool in order to perform those resummations.
 Results up to next-to-leading order for the ground state energy per particle of nuclear matter, the in-medium chiral quark condensate and pion self-energy are discussed.
\end{abstract}

\maketitle


\section{Introduction}
\label{sec:LacourA_intro}

An interesting achievement in nuclear physics would be the calculation of atomic nuclei and nuclear matter properties from microscopic inter-nucleon forces in a
systematic and controlled way.
This is a non-perturbative problem involving the strong interactions.
In the last decades, effective field theory (EFT) has been proven to be an indispensable tool to accomplish such an ambitious goal. 
In this work we employ chiral effective field theory ($\chi$EFT) \cite{LacourA_wein,LacourA_gl1} to nuclear systems \cite{LacourA_wein1,LacourA_wein2}, with nucleons and pions as the pertinent degrees of freedom.
For the lightest nuclear systems with two, three and four nucleons $\chi$EFT has been successfully applied \cite{LacourA_Epelbaum:2008ga}. 
For heavier nuclei one common procedure is to employ the chiral nucleon-nucleon (NN) potential derived in $\chi$EFT combined with standard many-body methods.
In \cite{LacourA_Oller:2009zt} we have derived a chiral power counting in nuclear matter that takes into account local multi-nucleon interactions simultaneously to the pion-nucleon interactions.
Many present applications to nuclei and nuclear matter only consider meson-baryon chiral Lagrangians without constraints from free NN scattering.
Others in fact include short range interactions, but loose contact to the vacuum by fitting free parameters to nuclear matter properties.
In \cite{LacourA_Lacour:2009ej} the techniques of unitary chiral perturbation theory (U$\chi$PT) \cite{LacourA_Oller:1998zr,LacourA_Oller:2000fj} have been applied to perform the resummation of the non-perturbative nature of the NN-interaction to vacuum phase shifts as well as to in-medium NN-scattering up to next-to-leading order (NLO).
Our novel power counting has been applied in \cite{LacourA_Oller:2009zt,LacourA_Lacour:2009ej,LacourA_Lacour:qc} to a variety of problems in nuclear matter.

\section{In-medium chiral power counting}
\label{sec:LacourA_cpc}

We consider the ground state of nuclear matter which, under the action of any time--dependent operator at asymptotic times, behaves as Fermi seas of nucleons \cite{LacourA_prcoller}. 
The Fermi seas are filled states up to a limiting Fermi momentum $\xi_\lambda=(3\pi^2\rho_\lambda)^{1/3}$, with $\rho_\lambda$ the density of the corresponding Fermi sea and $\lambda$ refers to the third component of the nucleon isospin.
The nucleon propagator in the medium contains a particle part propagating above the Fermi surface and a hole part propagating below \cite{LacourA_fetter}.
\begin{align}
 G(k)_\lambda&=\frac{\theta(|\vk|-\xi_\lambda)}{k^0-E_{\vk}+i\epsilon}+\frac{\theta(\xi_\lambda-|\vk|)}{k^0-E_{\vk}-i\epsilon}
=\frac{1}{k^0-E_{\vk}+i\epsilon}+2\pi i \, \delta(k^0-E_{\vk}) \, \theta(\xi_\lambda-|\vk|) ~.
 \label{eq:LacourA_nuc.pro}
\end{align}

In \cite{LacourA_prcoller} there has been established the concept of an ``in-medium generalized vertex'' (IGV).
Such type of vertices results because one can connect several bilinear vacuum vertices through the exchange of baryon propagators with the flow of a baryon number of one through a closed loop.
At least one Fermi sea insertion is needed, otherwise we would have a vacuum closed nucleon loop that in a low energy effective field theory is buried in the higher order counterterms.
In order to treat chiral Lagrangians with an arbitrary number of baryon fields (bilinear, quartic, etc.) we consider bilinear vertices like in \cite{LacourA_prcoller,LacourA_annp}, but allow the additional exchange of any type of heavy meson fields.
Those should be considered merely as auxiliary fields that allow one to find a tractable representation of the multi-nucleon interactions that result when the masses of the heavy mesons tend to infinity.
Consequently a heavy meson propagator is counted as ${\cal O}(p^0)$ due to their large masses.
In addition, we employ the non-standard counting case from the start \cite{LacourA_annp} and any nucleon propagator is considered as ${\cal O}(p^{-2})$.
In this way, no diagram whose chiral order is actually lower than expected by following the standard counting rules is lost.
We regard $m_\pi\sim \xi_\lambda\sim {\cal O}(p)$, which are considered much smaller than a hadronic scale $\Lambda_\chi$ of several hundred MeV that results by integrating out all other particle types, including nucleons with larger three-momentum, heavy mesons and nucleon isobars \cite{LacourA_wein2}.
The final formula for the chiral order $p^\nu$ of a given diagram is \cite{LacourA_Oller:2009zt}
\begin{equation}
\nu=4-E_\pi+\sum_{i=1}^{V_\pi}(n_i+\ell_i-4)+\sum_{i=1}^{V_B}(d_i+v_i+w_i-2)+V_\rho ~,
\label{eq:LacourA_fff}
\end{equation}
where $E_\pi$, $V_\pi$, $V_B$ and $V_\rho$ is the number of external pion lines, mesonic vertices, bilinear baryon vertices and IGVs, respectively. 
 The number of pion lines attached to the $i^{th}$ mesonic vertex is indicated by $n_i$ and $\ell_i$ corresponds to the chiral order of the latter.
 $d_i$ is the chiral order of the $i^{th}$ bilinear baryon vertex, $v_i$ is the number of all mesonic lines attached to it and $w_i$ the number of only the heavy lines.
 It is worth stressing that  $\nu$ is bound from below and  eq.\eqref{eq:LacourA_fff} implies conditions 
for augmenting the number of lines in a diagram without increasing the chiral power (see \cite{LacourA_Oller:2009zt} for details).
Because of the last term in eq.\eqref{eq:LacourA_fff} adding a new IGV to a connected diagram increases the counting at least by one unit.
The number $\nu$ given in eq.\eqref{eq:LacourA_fff} represents a lower limit for the actual chiral power $\mu$ of a diagram,
because in some instances (like standard $\chi$PT subdiagrams) the nucleon propagator may follow the standard counting.

For evaluations we employ the chiral effective Langrangian
\begin{equation}
  {\cal L}_{eff} = \sum_{n=1}^\infty{\cal L}_{\pi\pi}^{(2n)} + \sum_{n=1}^\infty{\cal L}_{\pi N}^{(n)} + \sum_{n=0}^\infty{\cal L}_{NN}^{(2n)} ~,
\end{equation}
where, from left to right one has the mesonic $\chi$PT Lagrangian, the Lagrangian of heavy baryon $\chi PT$ (HB$\chi$PT) \cite{LacourA_ulfrev} 
and the NN Lagrangian  \cite{LacourA_wein1,LacourA_wein2}.
The scattering amplitudes that follow these Lagrangians  can be found in \cite{LacourA_Lacour:2009ej}.
\section{Non-perturbative methods}
\label{sec:LacourA_nni}

\begin{figure}[ht]
\centerline{\epsfig{file=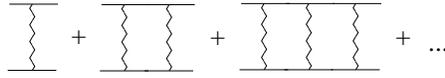,width=.4\textwidth,angle=0}}
\vspace{0.2cm}
\caption{\protect \small
Resummation of the two-nucleon reducible diagrams.
A wiggly line represents the interaction kernel $N_{JI}$ of fig.\ref{fig:LacourA_N}.
\label{fig:LacourA_sum}}
\end{figure} 

\begin{figure}[!ht]
  \centerline{
    \epsfig{file=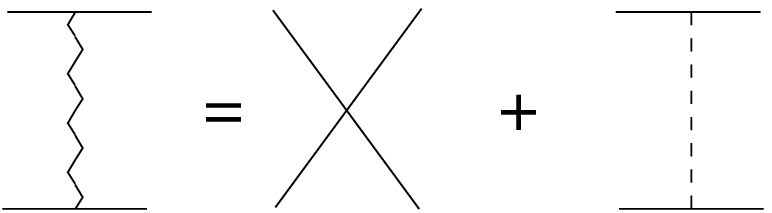,width=.24\textwidth,angle=0}
    ~~~~~~~~~~~~~
    \epsfig{file=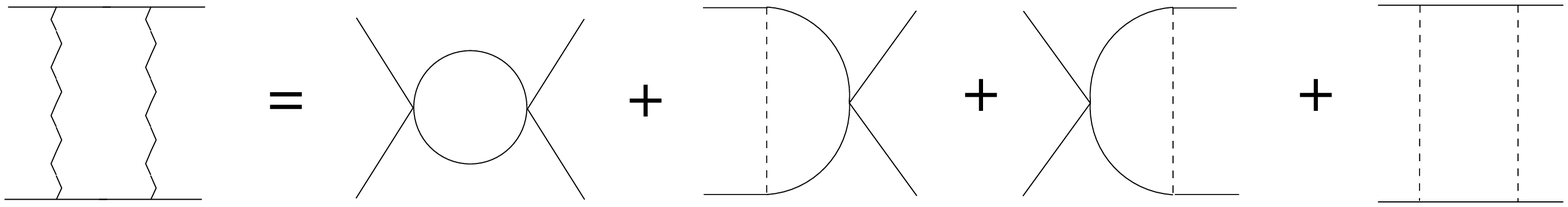,width=.5\textwidth,angle=0}
   }
  \caption{\protect \small
  The exchange of a wiggly line between two nucleons indicates the sum of the local and the one-pion exchange.
  A one-wiggle exchange corresponds to the leading order (LO), $L_{JI}^{(0)}$, and the
  two-wiggle exchange box diagram to NLO, $L_{JI}^{(1)}$.
  \label{fig:LacourA_N}}
\end{figure}

In \cite{LacourA_wein1,LacourA_wein2} it was argued that due to the large nucleon mass the two-nucleon reducible diagrams should be resummed, as it is schematically depicted in fig.\ref{fig:LacourA_sum}. This is also a consequence of eq.\eqref{eq:LacourA_fff} when considering lowest-order NN-interactions. 
The wiggly line is understood as a local NN interaction plus the exchange of a pion as depicted on the left hand side of fig.\ref{fig:LacourA_N}.


\begin{figure}[!ht]
  \psfrag{inf}{\tiny $\infty$}
  \psfrag{a}{\tiny $_{-\frac{m_\pi^2}{4}}$}
  \psfrag{C1}{\tiny $C_I$}
  \psfrag{C2}{\tiny $C_{II}$}
  \psfrag{e}{}
  \centerline{\epsfig{file=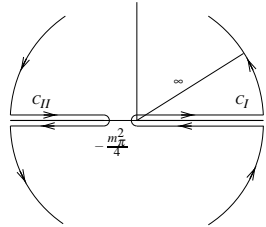,width=.23\textwidth,angle=0}}
  \caption{Right- and left-hand cuts of $T_{JI}(\ell,\bar{\ell},S)$.}
  \label{fig:cuts}
\end{figure}

We follow the techniques of U$\chi$PT \cite{LacourA_Lacour:2009ej,LacourA_Oller:1998zr,LacourA_Oller:2000fj} that performs the resummation partial wave by partial wave.
The relation for the scattering amplitude, $T_{JI}$, in U$\chi$PT is
\begin{equation}
  T_{JI}(\ell,\bar{\ell},S)=\left[I+ N_{JI}(\ell,\bar{\ell},S)\cdot g \right]^{-1}\cdot N_{JI}(\ell,\bar{\ell},S)~.
  \label{eq:LacourA_mucpt}
\end{equation}
It results by performing a once subtracted dispersion relation of the inverse of a partial wave amplitude taking an integration contour, denoted by $C_I\cup C_{II}$ in fig.\ref{fig:cuts}, closed by a circle with infinite radius that engulfs both the left- and the right-hand cut.
The quantum numbers are total isospin $I$, total spin $S$, total angular momentum $J$ and the out- and in-going orbital angular momenta $\ell,~\bar\ell$, in order.
The function $N_{JI}(A)$ is called the \textit{interaction kernel}.
The \textit{unitarity loop function} $g(A)$ shares the right-hand cut of the scattering amplitude and its discontinuity along this cut is $-im|\vk|/(2\pi)$.
In this way, the effects of the nucleon mass associated with the two-nucleon reducible diagrams are taken into account.
A once-subtracted dispersion relation can be written down given the degree of divergence for $\vk^2\rightarrow\infty$.
The integration contour is shown in fig.\ref{fig:cuts} by $C_I$. It results 
\begin{align}
 g(A) &= g(D)- \frac{m(A-D)}{4\pi^2}\int_0^\infty dk^2\frac{k}{(k^2-A-i\epsilon)(k^2-D)}= g_0 -i\frac{m\sqrt{A}}{4\pi} ~.
 \label{eq:LacourA_gA}
\end{align}
One subtraction has been taken at $D\leq 0$ so that the integral is convergent, yielding the subtraction constant $g_0$.
Note that $g(A)$ is real along the left-hand cut. 
According to eq.\eqref{eq:LacourA_mucpt} we can relate the imaginary part of the interaction kernel to the imaginary part of the scattering amplitude along the left-hand cut
\begin{align}
  \hbox{Im}N_{JI}&= |1+g N_{JI}|^2 \hbox{Im}T_{JI} \quad , \quad |\vk|^2<-\frac{m_\pi^2}{4}~.
\end{align}
We employ this result to write down a once-subtracted dispersion relation for $N_{JI}$.
The integration contour is shown in fig.\ref{fig:cuts} as $C_{II}$ and we obtain
\begin{align}
  N_{JI}(A) &= N_{JI}(D) + \frac{A-D}{\pi}\int_{-\infty}^{-m_\pi^2/4} dk^2\frac{\text{Im}T_{JI}(k^2)\,|1+g(k^2)N_{JI}(k^2)|^2}{(k^2-A-i\epsilon)(k^2-D)} ~.
  \label{eq:LacourA_N}
\end{align}
This is a non-linear  integral equation in $N_{JI}(A)$. However, one can obtain an approximate algebraic solution 
by choosing appropriately the subtraction constant $g_0$ in eq.\eqref{eq:LacourA_gA} so that $g(k^2)$ vanishes at 
one point in the low energy part of the left-hand cut. Making $g(-m_\pi^2)=0$ so that $|k|= i m_\pi$ one gets $g_0\sim -\frac{m m_\pi}{4\pi}=-0.54~m_\pi^2$, the 
natural size for $g_0$. 
In this way, with $g(k^2)$ suppressed in the low energy part of the left-hand cut, one can obtain a solution of eq.\eqref{eq:LacourA_N} as an expansion in powers of $g$. 
 For practical terms this is done as follows.  In a plain chiral calculation of NN partial waves the cuts are not resummed.
But we can match the chiral calculation with an expansion of the geometric series in eq.\eqref{eq:LacourA_mucpt}
\begin{equation}
  T_{JI}(\ell,\bar{\ell},S) = N_{JI}(\ell,\bar{\ell},S) - N_{JI}(\ell,\bar{\ell},S) \cdot g \cdot N_{JI}(\ell,\bar{\ell},S) + \ldots
  \label{eq:LacourA_T}
\end{equation}
up to the same number of two-nucleon reducible loops. 
In addition, one has the chiral expansion of the interaction kernel
\begin{equation}
  N_{JI} = \sum_{m=0}^\infty N_{JI}^{(m)} = N_{JI}^{(0)} + N_{JI}^{(1)} + \mathcal{O}(p^2) ~,
\end{equation}
where the chiral order is indicated by the superscript.
In order to extract $N_{JI}^{(n)}$, the ${\cal O}(p^n)$ chiral calculation of a NN partial wave comprising at most $n$ two-nucleon reducible loops is matched with eq.\eqref{eq:LacourA_T}
\begin{equation}
  N_{JI}^{(0)} + N_{JI}^{(1)} - N_{JI}^{(0)} \cdot g \cdot N_{JI}^{(0)} + \mathcal{O}(p^2) \doteq L_{JI}^{(0)} + L_{JI}^{(1)} + \mathcal{O}(p^2) ~.
  \label{eq:LacourA_match}
\end{equation}
On the right-hand-side of eq.\eqref{eq:LacourA_match} the chiral calculation is given, which is diagrammatically shown in fig.\ref{fig:LacourA_N}.
From eq.\eqref{eq:LacourA_match} we obtain $N_{JI}^{(0)}=L_{JI}^{(0)}$ and
\begin{equation}
  N_{JI}^{(1)} = L_{JI}^{(1)} + N_{JI}^{(0)} \cdot g \cdot N_{JI}^{(0)} ~,
\end{equation}
which corresponds to the difference between a full calculation of a two-nucleon reducible loop and the result obtained by factorizing the vertices on-shell, so that
 the right-hand cut is removed.
This is the reason why a two-nucleon reducible loop calculated in $\chi$PT is counted as ${\cal O}(p)$ in order to evaluate $N_{JI}$. 
This fact is incorporated in the interaction kernel $N_{JI}$, which can be improved order by order.

The evaluation of the NN scattering amplitudes in the nuclear medium can be obtained from the results in vacuum, since the only modification without increasing the chiral order corresponds to use the full in-medium nucleon propagator $G$, eq.\eqref{eq:LacourA_nuc.pro}.
This is directly accomplished by replacing
\begin{align}
 g \doteq L_{10,f} \quad\longrightarrow\quad L_{10} &= L_{10,f} + 2L_{10,m} + L_{10,d}
                                                     = L_{10,pp} + L_{10,hh}
 \label{eq:LacourA_L10}
\end{align}
in eq.\eqref{eq:LacourA_mucpt}, for any order of the NN scattering in the nuclear medium.
The subscripts $f$, $m$ and $d$ indicate zero, one or two Fermi sea insertions from the in-medium nucleon propagators.
The same process as discussed previously for the vacuum case is followed to fix $N_{JI}$, only that in all loop integrals the free nucleon propagators have to be exchanged by their full in-medium equivalents.

\section{Nuclear matter energy per particle}
\label{sec:LacourA_nmed}

\begin{figure}[!ht]
\psfrag{Vr=1}{\tiny $V_\rho=1$}
\psfrag{Vr=2}{\tiny $V_\rho=2$}
\psfrag{Op5}{\tiny ${\cal O}(p^5)$}
\psfrag{Op6}{\tiny ${\cal O}(p^6)$}
\centerline{\fbox{\epsfig{file=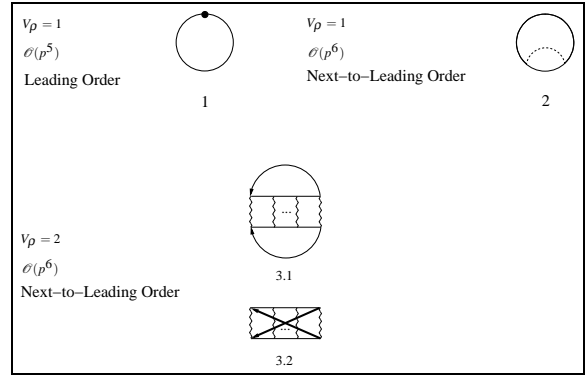,width=.5\textwidth,angle=0}}}
\caption{\protect \small
Contributions ${\cal E}_i$ to the nuclear matter energy density ${\cal E} = \langle\Omega|{\cal H}_{eff}|\Omega\rangle$ up to ${\cal O}(p^6)$.
Pions are depicted by dashed lines.
\label{fig:LacourA_all_nmed}}
\end{figure}

The diagrams required to study the problem of the nuclear matter ground state energy per particle $E/A={\cal E}/\rho$, by applying eq.\eqref{eq:LacourA_fff} up to NLO, are shown in fig.\ref{fig:LacourA_all_nmed}.
The kinetic energy closed by a Fermi sea insertion is depicted by diagram~1 and corresponds to the energy of a free Fermi gas.
All other diagrams are obtained by considering nucleon self-energy contributions which are closed with the full in-medium propagator $G$, eq.\eqref{eq:LacourA_nuc.pro}.
${\cal E}_3$ depends on $g_0$ both implicitly, due to the dependence   
of the partial wave amplitudes $T_{JI}$ on $g_0$, and explicitly in a linear manner.
For the latter dependence the symbol $\widetilde{g}_0$ was introduced in 
\cite{LacourA_Lacour:2009ej} (to which we refer for a more detailed discussion on 
$g_0$ and $\widetilde{g}_0$). 
For pure neutron matter both $g_0$ and $\widetilde{g}_0$ are obtained with the 
value $g_0=\widetilde{g}_0\simeq -0.6~m_\pi^2$, very close to their expected 
natural size~$g_0\simeq -0.54$~$m_\pi^2$. 
The case of symmetric nuclear matter requires some fine-tuning of $g_0$, with a 
final value of $g_0\simeq - m_\pi^2$, while $\widetilde{g}_0$ is kept on its 
natural value. 
\begin{figure}[!ht]
\centerline{\fbox{\epsfig{file=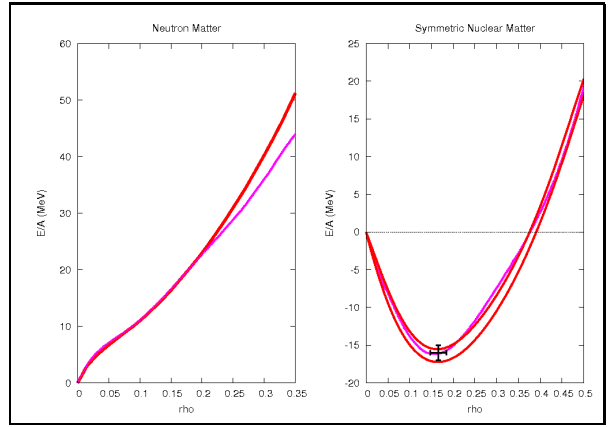,width=.52\textwidth,angle=0}}}
\caption{\protect \small
Energy per particle $E/A={\cal E}/\rho$ vs.\ the density $\rho$ for neutron (left panel) and symmetric nuclear matter (right panel).
\label{fig:LacourA_qc.en}}
\end{figure}
Results in perfect agreement with experiments were obtained for the saturation density, energy per particle and compression 
modulus \cite{LacourA_Lacour:2009ej}.
The equations of state for neutron (left panel) and symmetric nuclear matter (right panel) are shown in fig.\ref{fig:LacourA_qc.en}.
The magenta lines are from the many-body calculations of \cite{LacourA_urbana} employing to so-called realistic NN potentials.
The agreement between these references and our results is remarkable.

\section{In-medium chiral quark condensate}
\label{sec:LacourA_imcqc}

\begin{figure}[!ht]
\psfrag{Vr=1}{\tiny $V_\rho=1$}
\psfrag{Vr=2}{\tiny $V_\rho=2$}
\psfrag{Op5}{\tiny ${\cal O}(p^5)$}
\psfrag{Op6}{\tiny ${\cal O}(p^6)$}
\centerline{\fbox{\epsfig{file=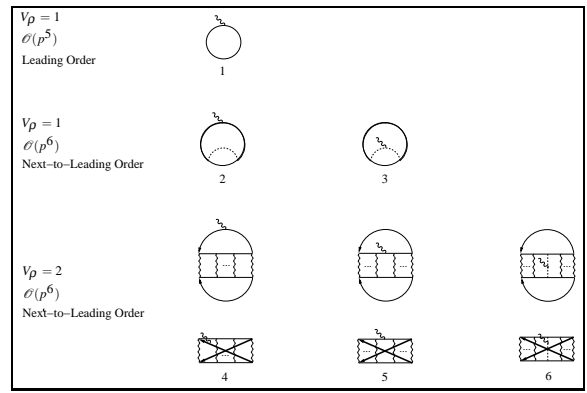,width=.5\textwidth,angle=0}}}
\caption{\protect \small
Contributions $\Xi_i$ to the in-medium chiral quark condensate $\langle\Omega | \bar{q}_i q_j | \Omega \rangle = - \delta/\delta s_{ij}(x){\cal Z}(v,a,s,p)|_{v,a,s,p=0}$  up to ${\cal O}(p^6)$.
The external scalar source in indicated by the wavy line.
\label{fig:LacourA_all_imcqc}}
\end{figure}
\begin{figure}[!ht]
\psfrag{k1}{\footnotesize $k_1$}
\psfrag{k2}{\footnotesize $k_2$}
\psfrag{k1-q}{\footnotesize $k_1-q$}
\psfrag{k2+q}{\footnotesize $k_2+q$}
\centerline{\epsfig{file=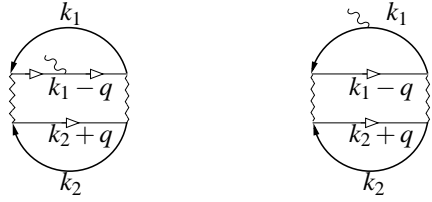,width=.38\textwidth,angle=0}}
\caption{\protect \small
One loop approximation in diagrams~5 (left) and 4 (right).
The pion scatters inside/outside the two-nucleon loop.
\label{fig:LacourA_cancel1loop}}
\end{figure}

The diagrams contributing to the problem of the in-medium chiral quark condensate, by applying eq.\eqref{eq:LacourA_fff} up to NLO, are shown in fig.\ref{fig:LacourA_all_imcqc}. Since there are no explicit quark degrees of freedom incorporated in the theory, the scalar chiral quark condensate is emulated by a scalar source \cite{LacourA_gl1}. Diagram~1 can be identified with the sigma-term of pion-nucleon scattering plus a small isospin breaking contribution.
Diagram~2 vanishes, while diagram~3 is suppressed by one order.
In diagram~6 the scalar source couples to a one-pion exchange of the NN interactions.
 For the diagrams 4 and 5 the scalar source is attached to the closing of lines in NN-scattering and to the nucleon propagators within the $NN$ reducible loop. 
It was found in \cite{LacourA_Lacour:qc} that those contributions cancel mutually.
The basic mechanism is depicted in fig.\ref{fig:LacourA_cancel1loop} in terms of a one-loop approximation.
By carefully considering the sums of all isospin and spin states one finds that both amplitudes just differ in the position of the derivative due to the squared nucleon propagators; for explicit expressions see \cite{LacourA_Lacour:qc}. After performing an integration by parts an exact cancellation of both contributions results.
The process of mutual cancellation can easily be generalized to any number of two-nucleon reducible loops in diagrams 4 and 5 in fig.\ref{fig:LacourA_all_imcqc}.
\begin{figure}[!ht]
\centerline{\epsfig{file=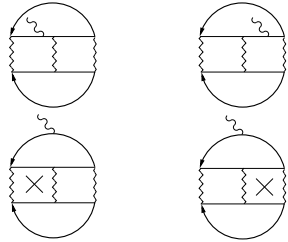,width=.25\textwidth,angle=0}}
\caption{\protect \small
The cross indicates the action of the derivative with respect to $k_1^0$.
In this way, the first diagram on the second row of the figure is the same as the one to the left of the first row but with opposite sign, such that they cancel each other.
The same applies to the second diagrams on the first and second rows.
\label{fig:LacourA_cancel2loop}}
\end{figure}
An $n+1$ iterated wiggly line exchange in these figures implies $n$ two-nucleon reducible loops.
The scalar source can be attached for $\Xi_5$ to any of them, while for $\Xi_4$ the derivative with respect to $k_1^0$, after performing the integration by parts, also acts on any of the iterated loops.
Hence,
\begin{align}
  \Xi_4+\Xi_5=0 ~.
\end{align}
Note, that this is a general argument following from the power counting and is not a property of the specific non-perturbative methods employed.

Putting together the results of all contributions arising from the topologies in fig.\ref{fig:LacourA_all_imcqc} the ratio between the in-medium and vacuum 
quark condensate to NLO reads
\begin{align}
\frac{\langle \Omega|\bar{q}_iq_j|\Omega\rangle}{\langle 0|\bar{q}_i q_j|0\rangle}
&=1-\frac{\sigma}{f_\pi^2 m_\pi^2}(\rho_p+\rho_n)
   +\frac{2 c_5 (\tau_3)_{ij}}{f_\pi^2}(\rho_p-\rho_n)
   -\frac{1}{f_\pi^2}\frac{\partial {\cal E}_3}{\partial m_\pi^2}~,
\label{eq:LacourA_ratio}
\end{align}
where the second and third term on the right hand side of eq.\eqref{eq:LacourA_ratio} rise from diagram~1 and the last term, which represents the quark mass derivative of the nuclear matter energy density, is identified with diagrams~3 and 6.
This result agrees with the Hellmann-Feynman theorem for nuclear matter.
Our findings, summarized in eq.\eqref{eq:LacourA_ratio}, actually explain observations made in earlier calculations \cite{LacourA_Kaiser:2007nv,LacourA_tubinguen} about the dominant role of the long-range part of  nucleon interactions for the in-medium chiral quark condensate.
 Due to the cancellation of diagrams~4 and 5 and the pion-mass squared derivative of ${\cal E}_3$ the pure the short-range part contributions vanish.

\section{In-medium pion self-energy}
\label{sec:LacourA_impse}

\begin{figure}[!ht]
\psfrag{Vr=1}{\tiny $V_\rho=1$}
\psfrag{Vr=2}{\tiny $V_\rho=2$}
\psfrag{Op4}{\tiny ${\cal O}(p^4)$}
\psfrag{Op5}{\tiny ${\cal O}(p^5)$}
\centerline{\fbox{\epsfig{file=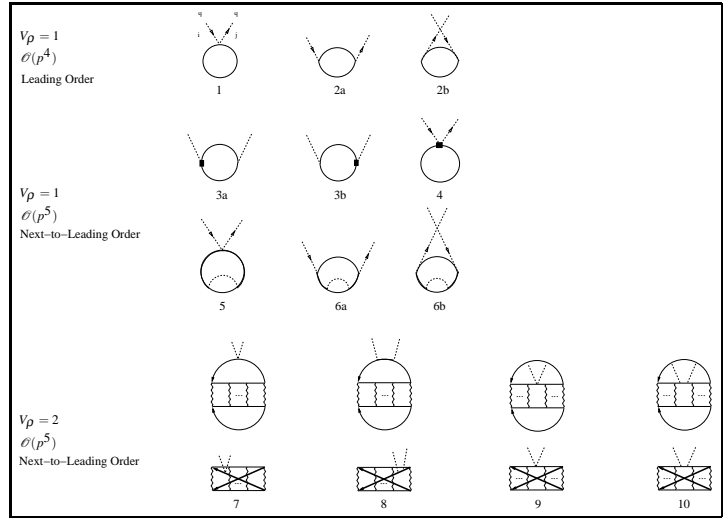,width=.625\textwidth,angle=0}}}
\caption{\protect \small
Contributions $\Pi_i$ to the in-medium pion self-energy up to ${\cal O}(p^5)$ entering the pion propagator $\Delta_\pi(q,\xi)=1/(q^2-m_\pi^2+\Pi(q,\xi)+i\epsilon)$ in the nuclear medium.
The squares correspond to NLO pion-nucleon vertices.
For diagrams~3a, 3b, 8 and 10 the pion lines can leave or enter the diagrams.
\label{fig:LacourA_all_impse}}
\end{figure}

Here, we apply the chiral counting given in eq.\eqref{eq:LacourA_fff} to calculate the pion self-energy in the nuclear medium up to NLO or ${\cal O}(p^5)$, with the different contributions shown in fig.\ref{fig:LacourA_all_impse}.
We only consider here those NLO contributions to the pion self-energy in the nuclear medium that involve the NN interactions, for the rest see \cite{LacourA_Oller:2009zt,LacourA_Lacour:2009ej}.
They are depicted in the diagrams of the last two rows in fig.\ref{fig:LacourA_all_impse}, where the ellipsis indicate the iteration of the two-nucleon reducible loops.
Since $V_\rho=2$ in these contributions one needs the NN scattering amplitude at ${\cal O}(p^0)$ to match with our required precision at NLO.
In the same way as described for the case of the in-medium chiral quark condensate, we find a mutual cancellation of the diagrams~7 and 9 and of the isovector parts of diagrams~8 and 10 \cite{LacourA_Oller:2009zt,LacourA_Lacour:2010zz}
\begin{align}
  \Pi_7+\Pi_9=0 ~, \quad \text{and} \quad \Pi_8^{iv}+\Pi_{10}^{iv}=0
\end{align}
Note that $\Pi_8+\Pi_{10}$ also contains isoscalar $1/m$ recoil corrections, which are higher order, that do not cancel and yield contributions from NN interactions.

\section{Conclusions}
\label{sec:LacourA_cno}

We have reviewed the development in \cite{LacourA_Oller:2009zt,LacourA_Lacour:2009ej,LacourA_Lacour:qc} of an effective field theory for nuclear matter that consistently treats local short-range and pion-mediated long-range multi-nucleon interactions consisting of a novel in-medium power counting and the non-perturbative methods known from U$\chi$PT.
We have calculated the nuclear matter energy per particle and found a naturally emerging saturation mechanism for symmetric nuclear matter and a weak parameter-dependence for neutron matter.
For the in-medium chiral quark condensate we found that our results obtained from calculations of the generating functional agrees with the Hellmann-Feynman theorem.
We also found the mechanism that explains why the short-range part of the NN-interaction contributes barely to the chiral quark condensate.
For the in-medium pion self-energy up to NLO we find that all contributions from NN-interactions mutually cancel, which explains why  pion-nucleon dynamics is phenomenologically successful for fitting pionic atom data.
Meanwhile it has also been shown in \cite{LacourA_Lacour:qc} that analogous cancellations happen for the calculation of the in-medium pion decay at NLO. Those findings show, that there are no contributions violating the Gell-Mann--Oakes--Renner relation at NLO in the nuclear medium.





\bibliographystyle{aipproc}   




\end{document}